# Pure Data and INScore: Animated notation for new music

Patricio F. Calatayud

Faculty member, UNAM

patricio.tics@gmail.com

Abstract: New music is made with computers, taking advantage of its graphics displays rather than its audio algorithms. Pure Data can be used to compose them. This essay will show a case study that uses Pure Data, in connection with INScore, for making a new type of score that uses animated notation or dynamic musicography for making music with performers. This sample was made by the author of the text, and it will show a number of notation possibilities that can be done using the combination of software. This will be accompanied by a simple prediction of what a musician could perform with it.

Keywords: Computer music, Animated notation, Pure Data, INScore.

## 1 Introduction

Music made with computers gained a change in direction in the last decades. Instead of only using their audio capabilities, many composers started to use their graphic power to compose music, generating scores that serves as a mediation between a creator and music performers (Hope 2017). Influenced by the Indeterminacy movement (e.g. *Calder's piece* by Earl Brown (1966)), many composers started to work on the expanded possibilities of connecting computer power, and the musician's gesture when performing dynamic musicography. The later concept refers to all drawings that are annotated in a music score, and have the intention to be gestures (i.e. are intended to have a musical utterance).

This connection could not be possible if all the available software were expensive, complex and subject to strict copyright laws. Fortunately, Pure Data (1997) is capable of working with complex algorithms to make music, and also it is open-free software. Although, in its beginnings it was designed to process audio information to obtain sound, there's always been the possibility of generating only data, and with the emergence of INScore (2010), we have the possibility of making music scores that can be transformed continuously, in real-time.

## 2 Animated notation and Dynamic musicography

Music made with computers normally takes advantage of audio algorithms and automated sequences for producing sound. Now we have a series of tools that uses the power of computers to mediate between a composer, and music performers: Animated notation or Dynamic musicography.

We must make a distinction. We can make music using computers for transcribing a score. Amongst those, are the ones that use the capabilities of computers to imitate what is outside of them (like score transcriptions of manuscripts or OMR digitalization). There are other music scores built inside the computer using their algorithm power (Arias et al. 2014, Oliver 2014). Much of those scores don't even need to be printed in paper, with its projection on a screen is enough to send musical information to a performer. The scores that we are presenting here are the ones that propose a change in music performance by changes in the presentation of notation, rather its complexity (see Hope 2017, Ross-Smith 2015). Dynamic musicography or animated notation transforms the score and the music notation with the aim of producing another type of musical gesture, and not necessarily a change in sound.

## 3 Case study

The basic notation material came from an old string quartet of the author of the text. A segmentation process took place, selecting a number of fragments that were transcribed again in MuseScore (2002), and exported as MIDI. This MIDI was imported in NoteAbilityPro (2000) for the purpose of generating a GUIDO notation file format (1998) that was appropriate for INScore, its graphics quality, and the possibility of modulating the information noted in the original fragments.

FIGURE 1 – Fragment of a string quartet extracted from its original context. Drawing made with MuseScore. Source: The author of the text (2019)

Also, for performing purposes we programmed four scrolling lines, each one in a different color, for the performers to follow. A line goes over a score fragment, as MuseScore does when it is getting audio from it. The idea is that each performer reads its part at his own time, also with the possibility of bending the scrolling line, getting another type of performance effect.

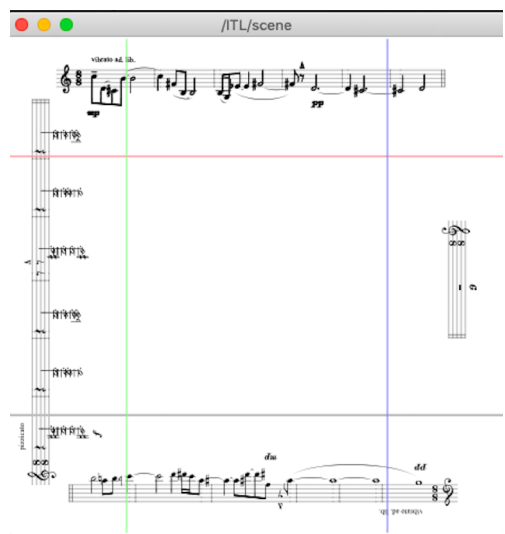

FIGURE 2 – Fragment of the new string quartet, displayed on the INScore scene. The image shows the four colored lines for each musician. Source: The author of the text (2019)

## 4 OSC implementation between Pure Data and INScore

Thanks to the enormous community that helps Pure Data users, we have a nurtured number of examples of OSC communications between any software and Pure Data. Although it is not easy to make the connection between the INScore software and Pure Data, when the two of them finally speak to each other, we can make transformations in the music notation, and let creativity flow.

The OSC messages formatting was initially a problem. Programming is not necessarily a composer's task -at least at this moment. So, the strategy was, as the reader may expect, dive into the many communities that Pure Data has. Unfortunately, when the case study was composed (2019), there were not many resources to find a patch that made the connection between the softwares; the documentation of INScore was implemented with a non-vanilla object -and its implications.

Fortunately, the Iglesias' MobMuPlat (2016) formatting subpatch did the work, and a connection was made successfully. A small problem came after with the float to integer conversion, but this was solved with the aid of the facebook Pure Data community.

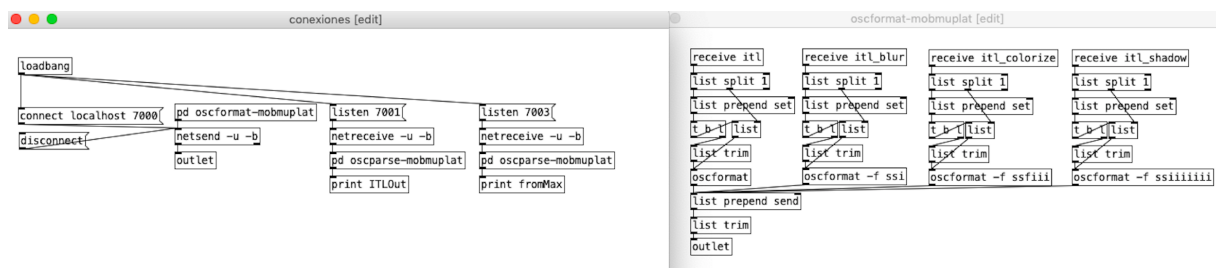

```
FIGURE 3 - Two examples of OSC patching in Pure Data.
```

## 5 Possibilities of new notation enhanced by Pure Data in combination with INScore

The listed possibilities below were made with Pure Data as the engine for sending continuous controller messages in OSC format to INScore, to be displayed on a screen. The information that modulates the fragments can be sent in all the possible strategies that digital musicians know: Algorithmic, statistical, stochastic, or any type of information strategy, and it can be performed in real-time or sequenced.

The results of the experimentation will be described as the INScore type of message and its implementation, followed by a way in which a musician can perform this feature.

a) The alpha channel gives the possibility of a "fade" effect on the score.

A crescendo or decrescendo is a possibility.

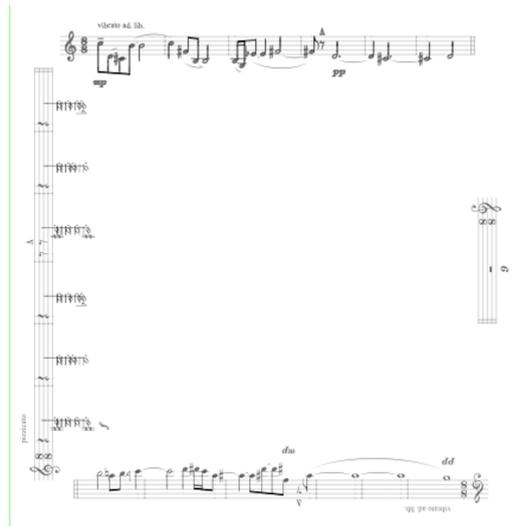

FIGURE 4 – Fragment of the new string quartet, displayed on the INScore scene. The score is fading in.

b) The X and Y planes can displace the fragments in the space made for the score.[1]

A movement in the X plane could be performed as change in the onset of the fragment, meaning a phasing texture, reflected between players.

Differentiations in the Y plane could be interpreted as Guido d'Arezzo understood: Changes in pitch. The fragment could be performed in a different interval (a kind of solmization) mapping a center location and its actual location.

c) Blurring or shadowing FX

Among all visual effects that we can use, INScore comes with a blurring one, useful to make distortion types of interpretation or, in the way it is used in the case study, a layering effect. There is also a similar effect called "shadow" that does mostly the same to the fragment. Also, we could get the opposite effect by changing the brightness or contrast of the fragment.

---

[1] In a mathematical way it is useful to distinguish between a transformation in the music notation, and the ones made on the space that they are displayed. The latter admits continuous transformations in the score as well.

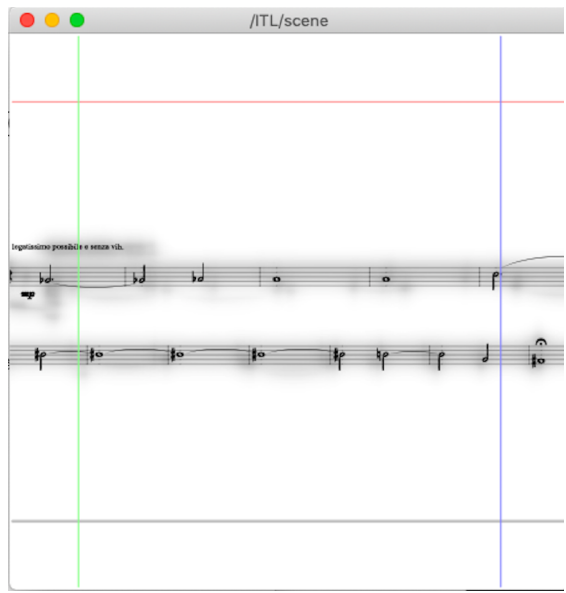

FIGURE 5 – Fragment of the new string quartet, displayed on the INScore scene. There are 4 fragments, three of them blurred.

d) Changes in color

Perhaps the most wanted effect in synesthesia simulations, like the ones made by Scriabin, the color of the notes or the fragment calls for a necessary dialog between composer and performer. It could be used for dynamic purposes, as in the work of Hoadley (2015).

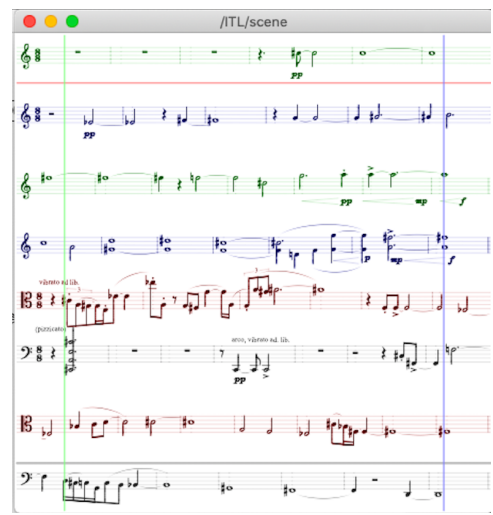

FIGURE 6 – Fragment of the new string quartet, displayed on the INScore scene. Fragments colored for each musician.

e) Scaling

Another effect that comes with INScore is a scaling factor that can be used for dynamic purposes, as in the *Jatekok* pieces by Gÿorgÿ Kurtag (1973-).

f) Angle can be done as a parameter of any individual fragment.

This is perhaps the most difficult one to perform. How could we perform differently the work *Mobile-stabile* of Silvano Bussotti (1959)?

Even in the times of the Augenmusik these types of angle transformations were performed as if they were in a normal position. A dialog between the performer and the composer is necessary. Again, the way that it is used in the case study is bending the arch position, accordingly.

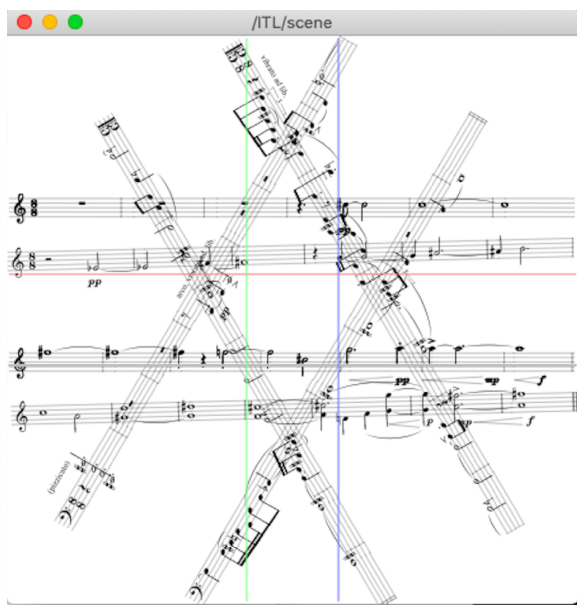

```
FIGURE 7 – Fragment of the new string quartet, displayed on the INScore
          scene. Fragments are displayed in a "star" angle.
```

g) Overlapping can be done with the X and Y planes.

The segmentation of the score is an arbitrary decision that can be approached by overlapping fragments and getting something different than a readability issue. Even a chord or an extension of a melody. This kind of transformation can be seen in the work of Mauricio Kagel's Prima Vista (1962-64).

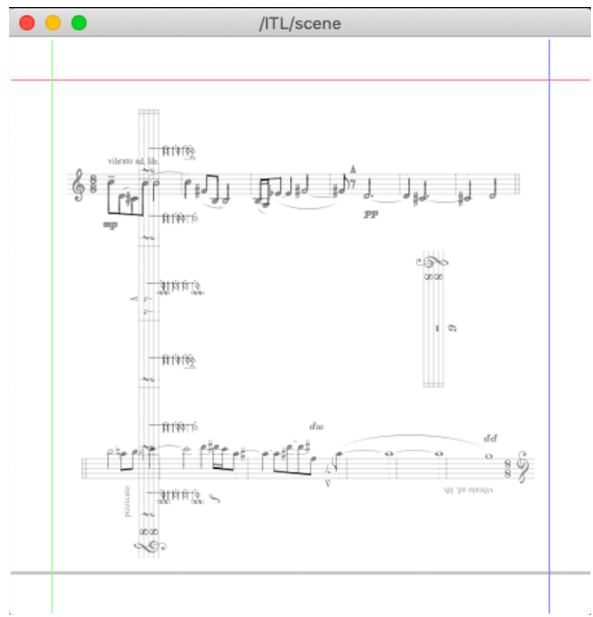

FIGURE 8 – Fragment of the new string quartet, displayed on the INScore scene. Fragments are overlapped.

There are also all other algorithmic possibilities of music display and material construction that are possible between Pure Data and OpenMusic (1998-), PGWL (2003) or like Oliver's implementation (2015).

On the other hand, these types of notation transformations are useful in a pedagogy field or even a research one. With the intent of demonstrating this, we take a small complement to the before study case, and present these "readability experiments", made with the purpose of asking the musicians what can be performed with a series of continuous transformations of notation, and its mathematical possibilities for analysis.

These exercises were made with the same structure as the one that we already explained, with the sole distinction of its target.

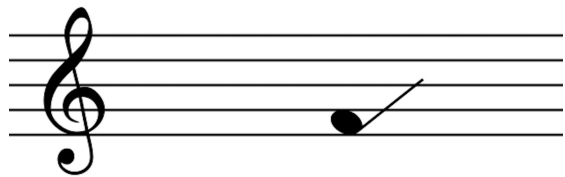

FIGURE 9 – Readability experiment "rotation a"..

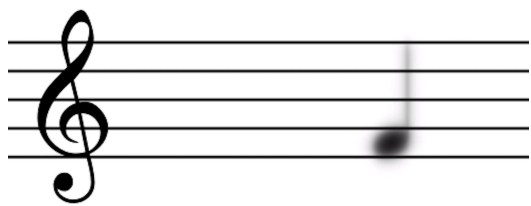

FIGURE 10 – Readability experiment "blurring".

## 6 Conclusions and future work

For some composers, fixed notation is a thing from the past. Each improvement in writing technologies changed the way we write. Exercises like those described above show that music performance has a way to expand its practices, and hopefully generate music in the process. These dynamic notations can also be used with visual art as iconographic material.

As this type of writing gradually becomes popular, many musicians will be setting a new ground of grammatical and orthographic rules that can be applied to this notation. Fortunately, Pure Data, with all its benefits, allows these strategies to come to take place in the music environment.

## 7 Acknowledgements

… for the intelligence of mathematics and cognition, needed for developing this project.